# Implication of Z-mediated FCNC on semileptonic decays $B_s \to \varphi l^+ l^-$ and $B^+ \to K^+ l^+ l^-$

P. Nayek[*], S. Biswas, P. Maji and S. Sahoo[**]

Department of Physics, National Institute of Technology Durgapur
Durgapur-713209, West Bengal, India
*E-mail: mom.nayek@gmail.com, **E-mail: sukadevsahoo@yahoo.com

**Abstract**
Rare B meson decays mediated by flavour changing neutral current (FCNC) transition play interesting role to probe the flavour sector of the standard model (SM). Generally at the tree level, FCNC processes are not allowed in the SM but occurs at the loop levels. This gives an excellent hunting ground for new physics (NP). From various experimental studies it is found that the FCNC processes having quark level transition $b \to s$ are challenging. Here, we investigate different kinematic observables like forward-backward asymmetry, differential branching ratio and lepton polarization asymmetry for semileptonic rare B decay modes $B_s \to \varphi l^+ l^-$ and $B^+ \to K^+ l^+ l^-$ ($l = \mu, \tau$) considering the contribution of Z-mediated FCNC. A noticeable deviation of the observables for these decay channels from the SM value is found because of non-universal $Z - bs$ coupling.

## 1. Introduction

The discovery of Higgs boson has left enough space for the standard model (SM) which gives an idea about the elementary particles and their strong, electromagnetic and weak interactions. The experiments of high energy physics are using two complementary approaches to solve some questions in the SM by introducing some new physics (NP) searches. The first approach is to study how the particles are smashing at the sufficiently large energy and producing different particles after collisions at the energy frontiers where ATLAS and CMS experiments at the LHC are the key representatives. The second one is at the borderland where Belle II and the LHCb experiment play an important role for the study of flavour physics. In the year 2018, a large data set of many kinematic observables on rare b-hadron decays are measured by the ATLAS, CMS and LHCb experiments at the LHC whereas Belle II experiments are started up. There are a number of anomalies between the theoretical and experimental values of flavour parameters. But no such direct evidence is found to undersatnd the effect of NP which can show the large discrepancy from the SM. There are some experimentally measured observables which gives small inconsistency from the SM are: $P_5'$, the angular observable [1-5] of $B \to K^* \mu^+ \mu^-$ decay channel, measurements of decay rate of $B_s \to \varphi \mu^+ \mu^-$ process with more than $3\sigma$ deviation [6], branching fraction of hadronic decays $b \to s\mu^+\mu^-$ [7, 8], measurement of lepton flavour universality (LFU) violating parameter $R_K = \mathcal{B}(B^+ \to K^+\mu^+\mu^-)/\mathcal{B}(B^+ \to K^+e^+e^-)$ [9-11] and $R_{K^*} = \mathcal{B}(B \to K^*\mu^+\mu^-)/\mathcal{B}(B \to K^*e^+e^-)$ [12, 13]. These inconsistencies introduce some anomalies in the rare B meson decays having the quark level transition $b \to s$. Generally, all neutral currents in the SM conserve flavour at the tree level and hence do not occur at the lowest order. But according to Glahsow-Iliopoulos-Maiani mechanism they are induced at the loop level [14]. Due to the dependency on weak mixing angles of the CKM matrix $V_{CKM}$, FCNC transitions are suppressed in the SM [15, 16]. Due to these two circumstances B meson decays induced by FCNC transitions are rare and a ground to test the SM and the possibilities of NP. FCNC mediated semileptonic rare B meson decays mediated by $b \to ql^+l^-$ quark level transition, (where $l = e, \mu, \tau$ and $q = s, d$) play the more promising role to modify the SM structure than the rare leptonic or radiative B decays since they are sensitive to NP contribution of the operators $O_7$, $O_9$ and $O_{10}$ through a broad set of parameters as lepton polarization asymmetry, branching ratio and forward-backward asymmetry. Experimentally it is shown that the semileptonic rare B meson decays mediated by quark level transition $b \to sl^+l^-$ are promising as their branching ratio ($\mathcal{O}(10^{-6})$ [17, 18]) is small. It is desired to know $B \to M$ form factors of the exclusive semileptonic decays $B \to Ml^+l^-$ (M is meson) over the full range $4m_l^2 < q^2 < (m_B - m_M)^2$. Due to intermediate u, c and t quarks in the matrix elements having the factors $V_{ub}V_{ud}^*$, $V_{cb}V_{cd}^*$ and $V_{tb}V_{td}^*$ the quark level transition process $b \to sl^+l^-$ have various contributions. Out of these three the first one is very small as compared to other. Due to unitary property of CKM matrix $V_{tb}V_{td}^* + V_{cb}V_{cd}^* \approx 0$ and the second factor is the negative of first one. Thus, the matrix element of this transition process $b \to sl^+l^-$ must contain only one independent CKM factor $V_{tb}V_{td}^*$ and this process is not precise to the CKM phases within the SM [19, 20]. So, we have to give a special attention on these semileptonic rare B meson decay channels [21, 22] both theoretically and experimentally. Moreover, the dileptons in these decay channels allow us to calculate many measurable quantities which can give a signal for NP. Now-a-days the experimental studies of the semileptonic decays induced by $b \to sl^+l^-$ have reached better position. Such a type of rare semileptonic B decay $B^+ \to K^+l^+l^-$ induced by quark level transition $b \to s$ are comparatively clean than the hadronic transitions and so it is expected that they are sensitive to some new interactions, such as 2HDM [23], supersymmetric theories [24], leptoquarks [25], non-universal $Z'$ model [26], single universal extra dimension [27] etc. Another type of such semileptonic rare B decays $B_s \to \varphi l^+l^-$ having quark level transition $b \to sl^+l^-$ are studied in various models, as scalar leptoquarks [28], supersymmetric [29], single universal extra dimension [30], non-universal Z' model and universal extra dimension model [31] and 331-Z' model [32].

Here, we are interested to study $B_s \to \varphi l^+l^-$ and $B^+ \to K^+l^+l^-$ ($l = \mu, \tau$) decays in non-universal Z model. There are various exotic scenarios where FCNC coupling of Z boson can be generated at the tree level e.g. (i) by including an extra $U(1)$ symmetry in the gauge group of the SM [33] and (ii) by adding the non-sequential generation of quarks [34]. In first case, due to the addition of extra gauge group $U(1)$ the FCNC coupling of $Z - Z'$ mixing will occur and the SM quarks remain family non-universal charges. In another approach, the pseudo CKM matrix needed to diagonalize the charged currents is no more unitary due to the addition of different numbers of up- and down type quarks. This leads to the FCNC coupling. These two possibilities (such as the extra $U(1)$ gauge group and the additional quarks) are connected in many extensions of the SM [35]. In this paper, we will follow the second approach where vector like down quark $D_4$ is added. This vector like down quark could mix with the SM three quark and give $4 \times 4$ mixing matrix K. For this mixing the charged current interactions remain unchanged except the case where $V_{CKM}$ is $3 \times 4$ upper submatrix of K. To gather the knowledge beyond the SM here we study the semileptonic rare B meson decay channels $B_s \to \varphi l^+ l^-$ and $B^+ \to K^+ l^+ l^-$ ($l = \mu, \tau$) in non-universal Z model.

This paper can be arranged as follows: The theoretical framework of effective Hamiltonian for $b \to s l^+ l^-$ transition in SM and definitions of some kinematic observables are presented briefly in Sec. 2. In Sec. 3, SM contributions on the variables associated with the semileptonic decay channel $B_s \to \varphi l^+ l^-$ are analized. In Sec. 4, $B^+ \to K^+ l^+ l^-$ decay in the SM is discussed. The contribution of Z-mediated FCNC to $B_s \to \varphi l^+ l^-$ and $B^+ \to K^+ l^+ l^-$ decay channels is discussed in Sec. 5,. Constraint on the coupling parameter $U_{sb}$ is studied using $B_s^0 - \bar{B}_s^0$ mixing data in Sec. 6. Our predicted values of parameters: forward backward asymmetry (FB), differential branching ratio and polarization asymmetry of $B_s \to \varphi l^+ l^-$ and $B^+ \to K^+ l^+ l^-$ decays are shown in Sec. 7 analytically and graphically. We conclude in Sec. 8.

## 2. Theoretical framework

To calculate the decay amplitude of the channels $B_s \to \varphi l^+ l^-$ and $B^+ \to K^+ l^+ l^-$ which have the basic quark level transition $b \to s l^+ l^-$ we have to follow some theoretical steps [30]:

(i) In the effective Hamiltonian we need a separation of long-distance effects which are present in matrix elements from short distance effects present in Wilson coefficients.

(ii) The matrix elements of the local quark bilinear operators $\langle \varphi | J | B_s \rangle$ can be calculated in terms of the form factors.

The semileptonic B meson decay channels $B_s \to \varphi l^+ l^-$ and $B^+ \to K^+ l^+ l^-$ have the quark level transition $b \to s l^+ l^-$. For this $b \to s$ transition the effective Hamiltonian can be presented as [31, 36-39]

$$H_{eff} = -\frac{4G_F}{\sqrt{2}} V_{tb} V_{ts}^* \sum_{i=1}^{10} C_i(\mu) O_i(\mu), \qquad (1)$$

where $O_i(\mu)$ ($i = 1, \dots \dots \dots 6$) are four-quark operators, $i = 7, 8$ are dipole operators and $i = 9, 10$ are semileptonic electroweak operators given in [40, 41]. $G_F$ is Fermi coupling constant and $C_i(\mu)$ are the Wilson coefficients calculated at the energy scale $\mu = m_b$. Here, we have considered the leading logarithmic approximation and taken the Wilson coefficients from [32]. The detail calculation of Wilson coefficients at the next-to-leading order (NLO) and also at the next-to-next-leading logarithm (NNLL) are done in [36, 40, 42-54]. The operators $\{O_i\}$ are given in [53, 54]. The unitary condition for the CKM matrix can be written as, $V_{tb} V_{ts}^* + V_{ub} V_{us}^* \approx -V_{cb} V_{cs}^*$ and the term $V_{ub} V_{us}^*$ can be safely ignored because $\frac{V_{ub} V_{us}^*}{V_{tb} V_{ts}^*} < 2 \times 10^{-2}$. The operators $O_7, O_9$ and $O_{10}$ are represented as follows

$$O_7 = \frac{e^2}{16\pi^2} m_b (\bar{s}\sigma_{\mu\nu} P_R b) F^{\mu\nu},$$

$$O_9 = \frac{e^2}{16\pi^2} (\bar{s}\gamma_\mu P_L b)(\bar{l}\gamma^\mu l),$$

$$O_{10} = \frac{e^2}{16\pi^2} (\bar{s}\gamma_\mu P_L b)(\bar{l}\gamma^\mu \gamma_5 l), \tag{2}$$

where $P_{L,R} = (1 \mp \gamma_5)/2$. The operator $O_{10}$ can not be induced by the insertion of the four quark operators. $C_{10}$ is independent of the energy scale as it does not renormalize under QCD corrections. Now the decay amplitude for this transition $b \to s l^+ l^-$ is written as

$$\mathcal{M} = \frac{G_F \alpha}{\sqrt{2}\pi} V_{tb} V_{ts}^* \Big\{ -2 C_7^{eff} \frac{m_b}{q^2} (\bar{s} i\sigma_{\mu\nu} q^\nu P_R b)(\bar{l}\gamma^\mu l) + C_9^{eff} (\bar{s}\gamma_\mu P_L b)(\bar{l}\gamma^\mu l) + C_{10}(\bar{s}\gamma_\mu P_L b)(\bar{l}\gamma^\mu \gamma_5 l) \Big\}. \tag{3}$$

The Wilson coefficient $C_9^{eff}$ corresponding to the operator $O_9$ has three parts and can be written as

$$C_9^{eff} = C_9^{SM}(\mu) + Y_{SD}(z, s') + Y_{LD}(z, s'), \tag{4}$$

where z and s' are denoted as $z = \frac{m_c}{m_b}$ and $s' = \frac{q^2}{m_b^2}$. The function $Y_{SD}(z, s')$ defines the short distance perturbative part that involves the indirect contributions from the matrix element of the four quark operators $\sum_{i=1}^{10} \langle l^+ l^- s | O_i | b \rangle$ and lies at the region far away from $c\bar{c}$ resonance regions. Now the short distance function is elaborately written as [36, 42]

$$Y_{SD}(z, s') = h(z, s')\big(3C_1(\mu) + C_2(\mu) + 3C_3(\mu) + C_4(\mu) + 3C_5(\mu) + C_6(\mu)\big)$$
$$- \frac{1}{2} h(1, s')\big(4C_3(\mu) + 4C_4(\mu) + 3C_5(\mu) + C_6(\mu)\big) - \frac{1}{2} h(0, s')(C_3 + 3C_4)$$
$$+ \frac{2}{9}(3C_3 + C_4 + 3C_5 + C_6), \tag{5}$$

with

$$h(z, s') = -\frac{8}{9} \ln(z) + \frac{8}{27} + \frac{4}{9} x$$
$$- \frac{2}{9}(2+x)\sqrt{|1-x|} \begin{cases} \ln\left|\frac{\sqrt{1-x}+1}{\sqrt{1-x}-1}\right| - i\pi, & x \equiv \frac{4z^2}{s'} < 1 \\ 2\arctan\frac{1}{\sqrt{x-1}}, & x \equiv \frac{4z^2}{s'} > 1 \end{cases} \tag{6}$$

and

$$h(0, s') = \frac{8}{27} - \frac{8}{9} \ln\frac{m_b}{\mu} - \frac{4}{9} \ln(s') + \frac{4}{9} i\pi. \tag{7}$$

Now another part of this Wilson coefficient $C_9^{eff}$ i.e. the long distribution contributions $Y_{LD}(z, s')$ have their origin in the $c\bar{c}$ intermediate states and is not obtained from the first principle of QCD. Basically they can be summerized in form of phenomenological Breit-Weigner formula. In our study, we have excluded this long distance resonance portion because it is far away from the part of our interest and experimental analysis also ignore this [30, 31, 55]. From eq. (3), differential decay rate of $B \to Ml^+l^-$ channel can be obtained as [56, 57]

$$\frac{d\Gamma(B \to Ml^+l^-)}{d\hat{s}dz} = \frac{m_B}{2^9\pi^3}\lambda^{1/2}(1,\hat{s},\widehat{m}_M{}^2)\sqrt{1-\frac{4\widehat{m}_l{}^2}{\hat{s}}}|\mathcal{M}|^2, \tag{8}$$

where $\hat{s} = \frac{s}{m_B{}^2}$, $\widehat{m}_l = \frac{m_l}{m_B}$ and $\widehat{m}_M = \frac{m_M}{m_B}$ are the dimensionless quantities. $\lambda(a,b,c) = a^2 + b^2 + c^2 - 2ab - 2ac - 2bc$ is the triangular function. $m_M$ is mass of meson M and $z = \cos\theta$. Here θ represents the angle between B three momenta and $l^-$ in the center of mass frame of the lepton pair. s denotes momentum transferred to the lepton pair. Now the FB asymmetry can be defined as [56-58]

$$A_{FB} = \frac{\int_0^1 dz \frac{d\Gamma}{d\hat{s}dz} - \int_{-1}^0 dz \frac{d\Gamma}{d\hat{s}dz}}{\int_0^1 dz \frac{d\Gamma}{d\hat{s}dz} + \int_{-1}^0 dz \frac{d\Gamma}{d\hat{s}dz}}. \tag{9}$$

To represent the expression of polarization asymmetries let us consider the unit vectors, S [56-60] to longitudinal (L), normal (N) and transverse direction (T) in the rest frame of $l^-$.

$$S_L^\mu \equiv (0, e_L) = \left(0, \frac{p_-}{|p_-|}\right)$$

$$S_N^\mu \equiv (0, e_N) = \left(0, \frac{q \times p_-}{|q \times p_-|}\right)$$

$$S_T^\mu \equiv (0, e_T) = (0, e_N \times e_L), \tag{10}$$

where $p_-$ and q are three momenta of $l^-$ and photon in CM frame of $l^+l^-$ system. Now longitudinal vector becomes

$$S_L^\mu = \left(\frac{|p_-|}{m_l}, \frac{E_-p_-}{m_l|p_-|}\right), \tag{11}$$

where other two remains same. So the polarization asymmetry is given by,

$$P_x(\hat{s}) \equiv \frac{\frac{d\Gamma(S_x)}{d\hat{s}} - \frac{d\Gamma(-S_x)}{d\hat{s}}}{\frac{d\Gamma(S_x)}{d\hat{s}} + \frac{d\Gamma(-S_x)}{d\hat{s}}}, \tag{12}$$

where $x = N, L, T$ represent normal, longitudinal and transverse polarization asymmetry respectively.

## 3. Standard model contribution on $B_s \to \varphi l^+ l^-$ decay

To study $B_s \to \varphi l^+ l^-$ decay theoretically, let us consider the decay matrix element between the initial and final states of meson. We have to first parameterize these decay matrix elements in terms of form factors. The form factors for $B_s \to \varphi$ transition are basically non-perturbative quantities. There are various approaches for calculating these form factors such as, light-cone sum rules, QCD sum rules, lattice QCD etc. In our study, we use light-cone sum rules to calculate these form factors related to the matrix element of $B_s \to \varphi l^+ l^-$ decay. The decay channel $B_s \to \varphi l^+ l^-$ includes the transition between the initial meson $B_s$ to final vector meson $\varphi$. To calculate decay amplitude of $B_s \to \varphi l^+ l^-$ at the hadron level, the free quark amplitudes between the initial and final states of meson is sandwiched. Thus hadronic matrix elements are

$$\langle \varphi(M_\varphi, p_2, \epsilon) | V_\mu(0) | B(M_B, p_1) \rangle, \qquad \langle \varphi(M_\varphi, p_2, \epsilon) | A_\mu(0) | B(M_B, p_1) \rangle$$

$$\langle \varphi(M_\varphi, p_2, \epsilon) | T_{\mu\nu}(0) | B(M_B, p_1) \rangle, \qquad \langle \varphi(M_\varphi, p_2, \epsilon) | T_{\mu\nu}(0) | B(M_B, p_1) \rangle. \qquad (13)$$

The form factors are shown in Appendix B. Using the form factors the decay matrix element can be given by [32, 61, 62]

$$\langle \varphi(M_\varphi, p_2, \epsilon) | V_\mu(0) | B(M_B, p_1) \rangle = 2g(q^2) \epsilon_{\mu\nu\alpha\beta} \epsilon^{*\nu} p_1^\alpha p_2^\beta, \qquad (14)$$

$$\langle \varphi(M_\varphi, p_2, \epsilon) | A_\mu(0) | B(M_B, p_1) \rangle = i\epsilon^{*\alpha} \big[ f(q^2) g_{\mu\alpha} + a_+(q^2) p_{1\alpha} P_\mu + a_-(q^2) p_{1\alpha} q_\mu \big], \qquad (15)$$

$$\langle \varphi(M_\varphi, p_2, \epsilon) | T_{\mu\nu}(0) | B(M_B, p_1) \rangle$$
$$= i\epsilon^{*\alpha} \big[ g_+(q^2) \epsilon_{\mu\nu\alpha\beta} P^\beta + g_-(q^2) \epsilon_{\mu\nu\alpha\beta} q^\beta + g_0(q^2) p_{1\alpha} \epsilon_{\mu\nu\beta\gamma} p_1^\beta p_2^\gamma \big], \qquad (16)$$

$$\langle \varphi(M_\varphi, p_2, \epsilon) | T^5_{\mu\nu}(0) | B(M_B, p_1) \rangle$$
$$= g_+(q^2) \big( \epsilon^*_\nu P_\mu - \epsilon^*_\mu P_\nu \big) + g_-(q^2) \big( \epsilon^*_\nu q_\mu - \epsilon^*_\mu q_\nu \big)$$
$$+ g_0(q^2) p_1 \epsilon^* \big( p_{1\nu} p_{2\mu} - p_{1\mu} p_{2\nu} \big), \qquad (17)$$

where $q = p_1 - p_2$, $P = p_1 + p_2$, vector $V_\mu = \bar{s}\gamma_\mu b$, tensor $T_{\mu\nu} = \bar{s}\sigma_{\mu\nu} b$ axial-vector $A_\mu = \bar{s}\gamma_\mu \gamma^5 b$ and pseudotensor $T^5_{\mu\nu}(0) = q\bar{\sigma}_{\mu\nu}\gamma_5 b$. We use the following notations: $\gamma^5 = i\gamma^0\gamma^1\gamma^2\gamma^3$, $\sigma_{\mu\nu} = \frac{i}{2}[\gamma_5, \gamma_\nu]$, $\epsilon^{0123} = -1$, $\gamma_5 \sigma_{\mu\nu} = -\frac{i}{2} \epsilon_{\mu\nu\alpha\beta} \sigma^{\alpha\beta}$ and $Sp(\gamma^5\gamma^\mu\gamma^\nu\gamma^\alpha\gamma^\beta) = 4i \epsilon^{\mu\nu\alpha\beta}$. Now we will discuss different physical observables in the SM.

# I. Differential decay rate (DDR)

The differential decay rate of the decay is represented as [32, 61]

$$\frac{d\Gamma(B_s \to \varphi l^+ l^-)}{d\hat{s}} = \frac{G_F^2 m_B^5 \alpha^2}{1536\pi^5} |V_{tb} V_{ts}^*|^2 \lambda^{\frac{1}{2}}(1,\hat{s},\hat{r}) \sqrt{1-\frac{4\hat{m}}{\hat{s}}} \left[\left(1+\frac{2\hat{m}}{\hat{s}}\right)\beta_V + 12\hat{m}\delta_V\right], \tag{18}$$

where

$$\beta_V = 2\lambda(1,\hat{s},\hat{r})\hat{s}|G(q^2)|^2 + \left[2\hat{s}+\frac{(1-\hat{r}-\hat{s})^2}{4\hat{r}}\right]|F(q^2)|^2 + \frac{\lambda^2(1,\hat{s},\hat{r})}{4\hat{r}}|H_+(q^2)|^2 - \frac{\lambda(1,\hat{s},\hat{r})}{2\hat{r}}(\hat{s}-1-\hat{r})R(q^2) \tag{19}$$

and

$$\delta_V = \frac{|C_{10}|^2}{2}\lambda(1,\hat{s},\hat{r})\left\{-2|g(q^2)M_B|^2 - \frac{3}{\lambda(1,\hat{s},\hat{r})}\right\}\left|\frac{f(q^2)}{M_B}\right|^2 + \frac{2(1+k)-\hat{s}}{4\hat{r}}|a_+(q^2)M_B|^2 + \frac{\hat{s}}{4\hat{r}}|a_-(q^2)M_B|^2 + \frac{1}{2\hat{r}}Re[f(q^2)a_+^*(q^2) + f(q^2)a_-^*(q^2)] + \frac{1-\hat{r}}{2\hat{r}}Re[M_B a_+(q^2)M_B a_-^*(q^2)], \tag{20}$$

with $r \equiv \left(\frac{M_\varphi}{M_B}\right)^2$, $\hat{m} \equiv \left(\frac{m_l}{M_B}\right)^2$ and $k = \lambda(s, m_b^2, m_s^2)/2\sqrt{s}$. Here, $m_s$ and $m_b$ are masses of the strange and bottom quark respectively. Now we can define the functions G, F, $H_+$ and R as follows

$$|G(q^2)|^2 = \left|C_9^{eff}(m_b, q^2)M_B g(q^2) - \frac{2C_7^{eff}(m_b)}{\hat{s}}\frac{m_b+m_s}{M_B}g_+(q^2)\right|^2 + |C_{10}(m_b)M_B g(q^2)|^2 \tag{21}$$

$$|F(q^2)|^2 = \left|C_9^{eff}(m_b, q^2)\frac{f(q^2)}{M_B} - \frac{2C_7^{eff}(m_b)}{\hat{s}}\frac{m_b-m_s}{M_B}(1-\hat{r})B_0(q^2)\right|^2 + \left|C_{10}(m_b)\frac{f(q^2)}{M_B}\right|^2 \tag{22}$$

$$|H_+(q^2)|^2 = \left|C_9^{eff}(m_b, q^2)M_B a_+(q^2) - \frac{2C_7^{eff}(m_b)}{\hat{s}}\frac{m_b-m_s}{M_B}B_+(q^2)\right|^2 + |C_{10}(m_b)M_B a_+(q^2)|^2 \tag{23}$$

$$R(q^2) = Re\left\{\left[C_9^{eff}(m_b,q^2)\frac{f(q^2)}{M_B} - \frac{2C_7^{eff}(m_b)}{\hat{s}}\frac{m_b - m_s}{M_B}(1-\hat{r})B_0(q^2)\right]\right.$$
$$\times \left[C_9^{eff}(m_b,q^2)M_B a_+(q^2) - \frac{2C_7^{eff}(m_b)}{\hat{s}}\frac{m_b - m_s}{M_B}(1-\hat{r})B_+(q^2)\right]^*\Big\}$$
$$+ |C_{10}(m_b)|^2 Re[a_+(q^2)f^*(q^2)]. \quad (24)$$

Another two functions $B_0(q^2)$ and $B_+(q^2)$ can be defined as

$$B_0(q^2) = g_+(q^2) + g_-(q^2)\frac{\hat{s}}{1-\hat{r}}, \quad (25)$$

and
$$B_+(q^2) = -\hat{s}M_B^2\frac{g_0}{2} - g_+(q^2). \quad (26)$$

From the eq. (18) the branching ratio of the decay channel $B_s \to \varphi l^+ l^-$ can be calculated.

## II. FB asymmetry

The FB asymmetry $A_{FB}$ can be represented as [32, 61]

$$A_{FB}(\hat{s}) = \frac{3\hat{s}\sqrt{1-4\widehat{m}/\hat{s}}\lambda^{\frac{1}{2}}(1,\hat{s},\hat{r})R_1(q^2)}{\left(1+\frac{2\widehat{m}}{\hat{s}}\right)\beta_V + 12\widehat{m}\delta_V}, \quad (27)$$

where

$$R_1(q^2) = Re\left\{\left[C_9^{eff}(m_b,q^2)M_B g(q^2) - \frac{2C_7^{eff}(m_b)}{\hat{s}}\frac{m_b + m_s}{M_B}g_+(q^2)\right]\right.$$
$$\times \left[C_{10}(m_b)\frac{f(q^2)}{M_B}\right]^*\Big\}$$
$$+ Re\left\{\left[C_9^{eff}(m_b,q^2)\frac{f(q^2)}{M_B} - \frac{2C_7^{eff}(m_b)}{\hat{s}}\frac{m_b - m_s}{M_B}(1-\hat{r})B_0(q^2)\right]\right.$$
$$\times [C_{10}(m_b)M_B g(q^2)]^*\Big\}. \quad (28)$$

The information about the NP and the sign of Wilson coefficients can be found from FB asymmetry. Finally, we determine another measurable quantity lepton polarization asymmetry for this decay channel.

## III. Polarization asymmetry

In [59, 60, 63-65] the importance of polarization asymmetry for various exclusive and inclusive semileptonic decay channels are explained briefly. The longitudinal polarization asymmetry is given by [32, 61],

$$P_L(\hat{s}) = \frac{2\sqrt{1-4\hat{m}/\hat{s}}}{\left(1+\frac{2\hat{m}}{\hat{s}}\right)\beta_V + 12\hat{m}\delta_V}$$

$$\times \left[2\lambda(1,\hat{s},\hat{r})\hat{s}R_G(q^2) + \left(2\hat{s} + \frac{(1-\hat{r}-\hat{s})^2}{4\hat{r}}\right)R_F(q^2) + \frac{\lambda^2(1,\hat{s},\hat{r})}{4\hat{r}}R_{H_+}(q^2)\right.$$

$$\left. - \frac{\lambda(1,\hat{s},\hat{r})}{4\hat{r}}(\hat{s}-1+\hat{r})R_R(q^2)\right], \tag{29}$$

where

$$R_G(q^2) = \text{Re}\left\{\left[C_9^{eff}(m_b,q^2)M_B g(q^2)\right.\right.$$
$$\left.\left. - \frac{2C_7^{eff}(m_b)}{\hat{s}}\frac{m_b+m_s}{M_B}g_+(q^2)\right][C_{10}(m_b)M_B g(q^2)]^*\right\}, \tag{30}$$

$$R_F(q^2) = \text{Re}\left\{\left[C_9^{eff}(m_b,q^2)\frac{f(q^2)}{M_B}\right.\right.$$
$$\left.\left. - \frac{2C_7^{eff}(m_b)}{\hat{s}}\frac{m_b-m_s}{M_B}(1-\hat{r})B_0(q^2)\right]\left[C_{10}(m_b)\frac{f(q^2)}{M_B}\right]^*\right\}, \tag{31}$$

$$R_{H_+}(q^2) = \text{Re}\left\{\left[C_9^{eff}(m_b,q^2)M_B a_+(q^2)\right.\right.$$
$$\left.\left. - \frac{2C_7^{eff}(m_b)}{\hat{s}}\frac{m_b-m_s}{M_B}B_+(q^2)\right][C_{10}(m_b)M_B a_+(q^2)]^*\right\}, \tag{32}$$

and

$$R_R(q^2) = \text{Re}\left\{\left[C_9^{eff}(m_b,q^2)\frac{f(q^2)}{M_B}\right.\right.$$
$$\left.\left. - \frac{2C_7^{eff}(m_b)}{\hat{s}}\frac{m_b-m_s}{M_B}(1-\hat{r})B_0(q^2)\right][C_{10}(m_b)M_B a_+(q^2)]^*\right\}$$
$$+ \text{Re}\left\{\left[C_9^{eff}(m_b,q^2)M_B a_+(q^2)\right.\right.$$
$$\left.\left. - \frac{2C_7^{eff}(m_b)}{\hat{s}}\frac{m_b-m_s}{M_B}(1-\hat{r})B_+(q^2)\right]\left[C_{10}(m_b)\frac{f(q^2)}{M_B}\right]^*\right\}. \tag{33}$$

## 4. Standard model contribution on $B^+ \to K^+ l^+ l^-$ decay

$B^+ \to K^+ l^+ l^-$ decay involves the $b \to s l^+ l^-$ quark level transition. The currents mainly responsible for semileptonic rare decay channel $B^+ \to K^+ l^+ l^-$ which is induced by $b \to s$ quark transitions are: vector current $V_\mu = \bar{s}\gamma_\mu b$, tensor current $T_{\mu\nu} = \bar{q}\sigma_{\mu\nu}b$ and pseudo-tensor current $T^5_{\mu\nu} = \bar{q}\sigma_{\mu\nu}\gamma_5 b$. The forward-backward asymmetry $A_{FB}$ for $B \to K l^+ l^-$ is very small within the SM [66] as the hadronic current for $B \to K$ transition have no axial-vector contribution. Four hadronic matrix elements are represented as [32, 61, 62]

$$\langle K(M_K,p_2,\epsilon)|\bar{s}\gamma_\mu b|B(M_B,p_1)\rangle, \qquad \langle K(M_K,p_2,\epsilon)|T_{\mu\nu}(0)|B(M_B,p_1)\rangle,$$

$$\langle K(M_K,p_2,\epsilon)|T^5_{\mu\nu}(0)|B(M_B,p_1)\rangle. \tag{34}$$

The matrix element of $B^+ \to K^+ l^+ l^-$ decay channel are represented as follows. In appendix C these form factors are broadly defined.

$$\langle K(M_K,p_2,\epsilon)|\bar{s}\gamma_\mu b|B(M_B,p_1)\rangle = f_+(q^2)P_\mu + f_-(q^2)q_\mu \tag{35}$$

$$\langle K(M_K,p_2,\epsilon)|T_{\mu\nu}(0)|B(M_B,p_1)\rangle = -2it(q^2)(p_{1\mu}p_{2\nu} - p_{1\nu}p_{2\mu}) \tag{36}$$

$$\langle K(M_K,p_2,\epsilon)|T^5_{\mu\nu}(0)|B(M_B,p_1)\rangle = t(q^2)\epsilon_{\mu\nu\alpha\beta}P^\alpha q^\beta \tag{37}$$

### I. Differential decay rate (DDR)

Differential decay rate is represented as [32, 61]

$$\frac{d\Gamma(B^+ \to K^+ l^+ l^-)}{d\hat{s}} = \frac{G_F^2 m_B^5 \alpha^2}{1536\pi^5}|V_{tb}V_{ts}^*|^2 \lambda^{\frac{1}{2}}(1,\hat{s},\hat{r})\sqrt{1-\frac{4\hat{m}}{\hat{s}}}\left[\left(1+\frac{2\hat{m}}{\hat{s}}\right)\lambda(1,\hat{s},\hat{r})\beta_P + 12\hat{m}\delta_P\right], \tag{38}$$

where

$$\beta_P = \left|C_9^{eff}(m_b,q^2)f_+(q^2) + 2(m_b+m_s)C_7^{eff}(m_b)t(q^2)\right|^2 + |C_{10}(m_b)f_+(q^2)|^2 \tag{39}$$

and

$$\delta_P = |C_{10}|^2\left\{\left(1+\hat{r}-\frac{\hat{s}}{2}\right)|f_+(q^2)|^2 + (1-\hat{r})Re[f_+(q^2)f_-^*(q^2)] + \frac{\hat{s}}{2}|f_-(q^2)|^2\right\}. \tag{40}$$

Here, we have used $r \equiv \left(\frac{M_K}{M_B}\right)^2$ and $\hat{m} \equiv \left(\frac{m_l}{M_B}\right)^2$. Using the eq. (38) the differential branching ratio of this decay mode can be calculated.

### II. Polarization asymmetry

Finally, we will calculate the longitudinal polarization asymmetry for this rare semileptonic decay channel $B^+ \to K^+ l^+ l^-$. The longitudinal polarization asymmetry is as follows [32, 61]

$$P_L(\hat{s}) = \frac{2\sqrt{1-4\hat{m}/\hat{s}}\lambda(1,\hat{s},\hat{r})}{\left(1+\frac{2\hat{m}}{\hat{s}}\right)\lambda(1,\hat{s},\hat{r})\beta_P + 12\hat{m}\delta_P}Re\{[C_9^{eff}(m_b,q^2)f_+(q^2) + 2(m_b+m_s)C_7^{eff}(m_b)t(q^2)]C_{10}^*f_+^*(q^2)\}. \tag{41}$$

## 5. Contribution of Z-mediated FCNC on $B_s \to \varphi l^+ l^-$ and $B^+ \to K^+ l^+ l^-$ decay modes

We consider a NP model with an extended matter due to the addition of extra vector like down quark $D_4$. Basically iso-singlet quarks are introduced in various extensions of the SM such as the low energy limit of $E_6$ GUT models [67]. It is needed to study $3 \times 3$ CKM matrix with the deviations of unitary constraint and this deviation basically arises due to the mixing between the singlet type down quark and three SM quarks. The eigenstates of down sector are involved with the up sector interaction eigenstates by a $4 \times 4$ unitary matrix K. Now the charged current interaction is given by [34]

$$\mathcal{L}_{int}^{W} = \frac{g}{\sqrt{2}} \left( W_\mu^{-1} J^{\mu^+} + W_\mu^{+1} J^{\mu^-} \right), \tag{42}$$

with

$$J^{\mu^-} = V_{ij} \bar{u}_{iL} \gamma^\mu d_{jL}. \tag{43}$$

Here V, the charged-current mixing matrix is a $3 \times 4$ submatrix of unitary matrix K:
$$V_{ij} = K_{ij} \quad \text{for } i = 1,2,3 \text{ and } j = 1,2,3,4 \tag{44}$$

The neutral current interaction is represented as [67-84],
$$\mathcal{L}_Z = \frac{g}{\cos\theta_w} Z_\mu \left( J^{\mu 3} - \sin^2\theta_w J_{em}^\mu \right) \tag{45}$$

where

$$J^{\mu 3} = -\frac{1}{2} U_{\alpha\beta} \bar{d}_{\alpha L} \gamma^\mu d_{\beta L} + \frac{1}{2} \delta_{ij} \bar{u}_{iL} \gamma^\mu u_{jL}. \tag{46}$$

so,

$$\mathcal{L}_Z = \frac{g}{2\cos\theta_w} Z_\mu \left[ \bar{u}_{iL} \gamma^\mu u_{iL} - U_{\alpha\beta} \bar{d}_{\alpha L} \gamma^\mu d_{\beta L} - 2\sin^2\theta_w J_{em}^\mu \right], \tag{47}$$

with

$$U_{\alpha\beta} = \sum_{i=u,c,t} V_{\alpha i}^\dagger V_{i\beta}. \tag{48}$$

Here, U denotes the neutral-current mixing matrix for the down quark sector. As V is not unitary so also $U \neq 1$. Basically the non-diagonal elements are not vanished and so

$$U_{\alpha\beta} = -K_{4\alpha}^* K_{4\beta} \neq 0, \text{for } \alpha \neq \beta. \tag{49}$$

As these $U_{\alpha\beta}$ are nonvanishing, they would give a signal for NP. This can revise various pedictions of SM for the FCNC processes. Basically in two different ways NP effects in non-universal Z model can arise: either changing the SM structure of effective Hamiltonian or by adding new terms in Wilson coefficients. Here, we are interested to modify the Wilson coefficients $C_9^{eff}$ and $C_{10}$. Now the corresponding effective Hamiltonian in this model can be given by [67-70]

$$\mathcal{H}_{eff} = \frac{G_F}{\sqrt{2}} U_{sb} [\bar{s} \gamma^\mu (1-\gamma_5) b] [\bar{l} (C_V^l \gamma_\mu - C_A^l \gamma_\mu \gamma_5) l], \tag{50}$$

here $C_A^l$ and $C_V^l$ are axial vector and vector couplings of lepton sector with Z boson i.e. $Zl^+l^-$. These couplings are given as

$$C_V^l = -\frac{1}{2} + 2\sin^2\theta_w, \qquad C_A^l = -\frac{1}{2}. \tag{51}$$

The effective Hamiltonian in this model described by eq. (50) is same as that of the SM, like $\sim (V-A) \times (V-A)$ form. So the contribution of Z boson on the four quark operators, semileptonic electroweak operators and dipole operators will be same as that of the SM and it only modify the Wilson coefficients $C_9^{eff}$ and $C_{10}$. The total contributions on these two Wilson coefficients $C_9$ and $C_{10}$ can be written as

$$C_9^{Total} = C_9^{eff} + C_9^{NP}, \tag{52}$$

$$C_{10}^{Total} = C_{10} - C_{10}^{NP}, \tag{53}$$

with

$$C_9^{NP} = \frac{2\pi}{\alpha} \frac{U_{sb} C_V^l}{V_{tb} V_{ts}^*}, \tag{54}$$

$$C_{10}^{NP} = \frac{2\pi}{\alpha} \frac{U_{sb} C_A^l}{V_{tb} V_{ts}^*}. \tag{55}$$

The coupling $U_{sb}$ representing the Z-b-s strength is a complex quantity and is parameterized as $U_{sb} = |U_{sb}|e^{i\varphi_{sb}}$ and it induces $\varphi_{sb}$ which is the weak phase difference between the contributions of NP and the SM. The values of $C_9$ and $C_{10}$ are opposite to each other in the SM and also in the NP contributions, so one can get destructive and constructive interference of the SM and NP amplitude for $\varphi_{sb} =$ zero and $\pi$ respectively. Now, we need to know the constraints on the FCNC coupling of the Z boson to the quark sector i. e. $U_{sb}$ that can be obtained from $B_s^0 - \bar{B}_s^0$ mixing parameters which is discussed in the next section.

## 6. Constraint on quark coupling from $B_s^0 - \bar{B}_s^0$ mixing

The mass difference between the mass eigen states denoted as $(\Delta M_s)$ of $B_s$ meson characterizes $B_s^0 - \bar{B}_s^0$ mixing. In the SM, $B_s^0 - \bar{B}_s^0$ mixing arises through one-loop level box diagram with top quark and W boson in the loop and hence it is precise to NP contributions. The effective Hamiltonian representing the $|\Delta B = 2|$ transitions, induced by the box diagram in the SM, can be written as [85, 86]

$$\mathcal{H}_{eff}^{SM}(\Delta B = 2) = \frac{G_F^2}{16\pi^2} M_W^2 (V_{tb} V_{ts}^*)^2 C^{LL}(\mu_b) \mathcal{O}^{LL} + h.c., \tag{56}$$

where the operator $\mathcal{O}^{LL}$ is defined as [68]

$$\mathcal{O}^{LL} = [\bar{s}\gamma_\mu(1-\gamma_5)b][\bar{s}\gamma^\mu(1-\gamma_5)b], \tag{57}$$

and $C^{LL}$ is the corresponding loop function. Now the $B_s^0 - \bar{B}_s^0$ mixing amplitude in the SM is represeted as [85, 87]

$$M_{12}^{SM}(s) = \frac{1}{2M_{B_s}} \langle B_s^0 | \mathcal{H}_{eff}^{SM}(\Delta B = 2) | \bar{B}_s^0 \rangle$$

$$= \frac{G_F^2}{12\pi^2} M_W^2 (V_{tb}V_{ts}^*)^2 (\hat{B}_{B_s} f_{B_s}^2) M_{B_s} \eta_B S_0(x_t) [\alpha_s(\mu_b)]^{-\frac{\gamma_Q^{(0)}}{2\beta_0}} \left[1 + \frac{\alpha_s(\mu_b)}{4\pi} J_5\right],$$
(58)

to evaluate the matrix element we have used the vacuum insertion method and we get

$$\langle B_s^0 | O^{LL} | \bar{B}_s^0 \rangle = \frac{8}{3} \hat{B}_{B_s} f_{B_s}^2 M_{B_s}^2.$$
(59)

Here $\hat{B}_{B_s}$ is the bag parameter, $x_t = \left(\frac{m_t}{M_W}\right)^2$ and the "Inami-Lim" loop function $S_0(x_t)$ which can be written as

$$S_0(x_t) = \frac{4x_t - 11x_t^2 + x_t^3}{4(1-x_t)^2} - \frac{3x_t^3 \ln x_t}{2(1-x_t)^3},$$
(60)

the parameters used in eq. (58) have values $\gamma_Q^{(0)} = 4$, $\beta_0 = \frac{23}{3}$ and $J_5 = 1.627$ [69]. The mass difference between the heavy and light mass eigenstates which describes the strength of $B_s^0 - \bar{B}_s^0$ mixing is related to mixing amplitude as $\Delta M_s = 2|M_{12}^{SM}|$. The mass difference $\Delta M_s$ in the SM is calculated by taking all the particle masses from [88], $\eta_B = 0.551$, the bag parameter $\hat{B}_{B_s} = 1.320 \pm 0.017 \pm 0.03$ from [89] as

$$\Delta M_s^{SM} = 17.426 \pm 1.057 \, ps^{-1} \, [90, 91].$$
(61)

The experimental value of $\Delta M_s$ is [88]

$$\Delta M_s = 17.757 \pm 0.02 \, (stat.) \pm 0.007 (syst.).$$
(62)

The ratio of the experimental and SM values found to be

$$\Delta M_s / \Delta M_s^{SM} = 1.018 \pm 0.062$$
(63)

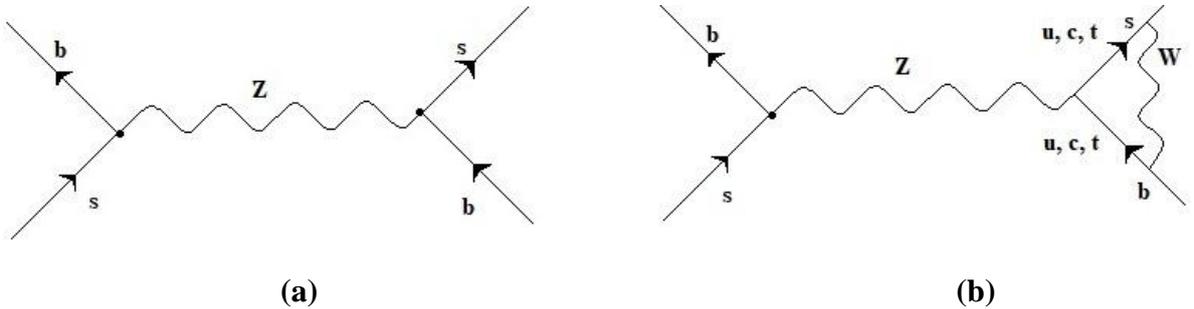

**Fig. 1.** Feynman diagrams for $B_s^0 - \bar{B}_s^0$ mixing in vector like down quark model. Here, blobs represent tree level flavour changing vertex [92].

In this model, there are two extra contributions to the $B_s^0 - \bar{B}_s^0$ mixing amplitude [92]. First contribution is due to the tree level FCNC mediated Z bososn with two non-standard Z-b-s coupling (Fig. 1(a)). In another contribution one non-standard Z-b-s coupling as well as one

SM loop-induced Z-b-s coupling are also present shown in Fig. 1(b). The effective FCNC mediated Lagrangian in this model can be written as [92]

$$\mathcal{L}^Z_{FCNC} = -\frac{g}{2\cos\theta_w} U_{sb} \bar{s}_L \gamma^\mu b_L Z_\mu. \tag{64}$$

Here $U_{sb}$ is the coupling of Z boson to the quark sector. The structure of effective Hamiltonian which is induced by tree level FCNC mediated Z boson is given by

$$\mathcal{H}^Z_{eff} = \frac{G_F}{\sqrt{2}} U_{sb}^2 \eta_Z (\bar{s}_L \gamma^\mu b_L)(\bar{s}_L \gamma_\mu b_L), \tag{65}$$

where $\eta_Z = (\alpha_s(\mu_Z))^{6/23}$ represents the QCD correction factor. From effective Hamiltonian the matrix element can be defined as

$$M_{12}^Z = \frac{G_F}{3\sqrt{2}} U_{sb}^2 \eta_Z \hat{B}_{B_s} f_{B_s}^2 M_{B_s}. \tag{66}$$

Now considering the second contribution the matrix element can be defined as

$$M_{12}^{SM+Z} = \frac{G_F^2}{3\pi^2} (V_{tb} V_{ts}^*) U_{sb} \eta_{Zt} M_W^2 C_0(x_t) \hat{B}_{B_s} f_{B_s}^2 M_{B_s}, \tag{67}$$

where

$$C_0(x_t) = \frac{x_t}{8}\left(\frac{x_t - 6}{x_t - 1} + \frac{3x_t + 2}{(x_t - 1)^2}\log x_t\right). \tag{68}$$

Now the total contributions of both SM and Z model to the mass difference is given by

$$\Delta M_s = \Delta M_s^{SM} + \Delta M_s^Z + \Delta M_s^{SM+Z}$$

$$= \Delta M_s^{SM} \left| 1 + \frac{4\pi^2 |U_{sb}|^2 e^{2i\varphi_{sb}} (\alpha_s(\mu_Z))^{6/23}}{G_F \sqrt{2} M_W^2 (V_{tb} V_{ts}^*)^2 \eta_B S_0(x_t)[\alpha_s(\mu_b)]^{-\frac{\gamma_Q^{(0)}}{2\beta_0}}\left[1 + \frac{\alpha_s(\mu_b)}{4\pi} J_5\right]} + \frac{4|U_{sb}|e^{2i\varphi_{sb}} C_0(x_t)}{(V_{tb} V_{ts}^*) S_0(x_t)[\alpha_s(\mu_b)]^{-\frac{\gamma_Q^{(0)}}{2\beta_0}}\left[1 + \frac{\alpha_s(\mu_b)}{4\pi} J_5\right]}\right|.$$

(69)

In the above expression the CKM parameter $V_{tb}V_{ts}^* = -|V_{tb}V_{ts}|e^{i\beta_s}$, where $\beta_s$ is the phase of $V_{ts} = -|V_{ts}|e^{-i\beta_s}$ and $\beta_s = -1.1°$. We have taken all the required parameters from [88] and calculate the ratio of $\Delta M_s$ to $\Delta M_s^{SM}$ in terms of $U_{sb}$, the coupling parameter and new weak phase angle $\varphi_{sb}$. Now by varying $\Delta M_s/\Delta M_s^{SM}$ within its 2σ range the allowed space of the new weak phase and the coupling parameter can be calculated from $U_{sb} - \varphi_{sb}$ plane which is shown in Fig. 2.

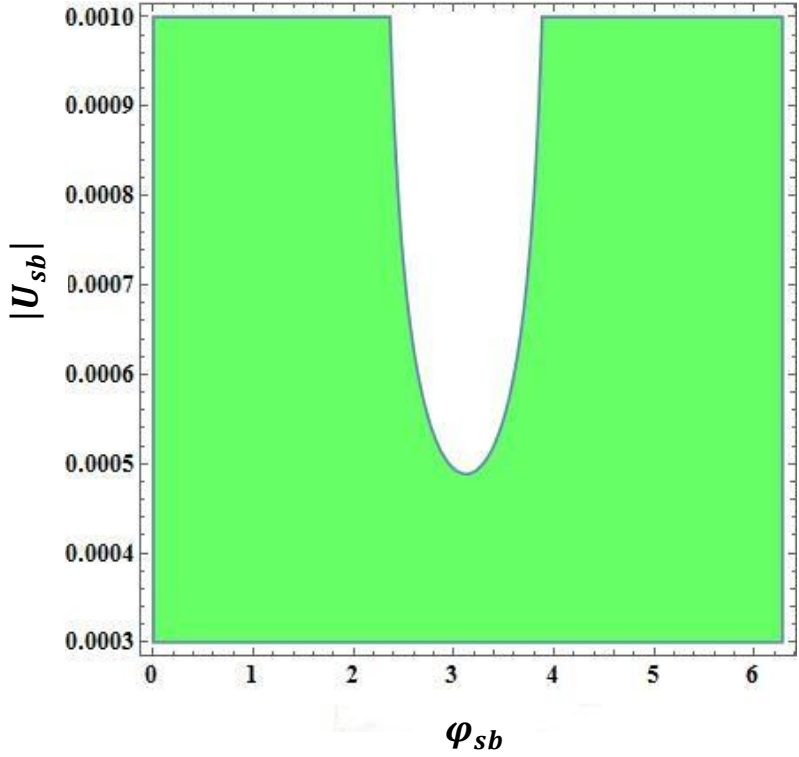

**Fig. 2.** Allowed space of $|U_{sb}|$ and $\varphi_{sb}$ (degree) obtained from $B_s^0 - \bar{B}_s^0$ mixing

In the above figure we can observe that the phase $\varphi_{sb}$ is tightly constrained for the higher value of $|U_{sb}|$. It indicates that for $|U_{sb}| \leq 0.00048$, the total range of $\varphi_{sb}$ is allowed i. e. from 0 to $2\pi$. The constraint on $|U_{sb}|$ can be obtained from $B \to X_S l^+ l^-$ i.e. $4.89 \times 10^{-4}$ which is consistent with the value calculated from graphical representation of $B_s^0 - \bar{B}_s^0$ mixing.

## 7. Numerical Analysis

We have taken all required parameters from [88] shown in Table 2 of Appendix A. From the above discussions, we have taken $|U_{sb}| = 0.00048$. The weak phase difference can be considered as 0 for destructive interference and $\pi$ for constructive interference between SM and NP amplitude. Considering eqs. (52) and (53), we show the variations of kinematic observables graphically for $B_s \to \varphi l^+ l^-$ and $B^+ \to K^+ l^+ l^-$ decay channels with $q^2$ and new weak phase in this section.

In Fig. 3, fixing the value of coupling parameter as $|U_{sb}| = 0.00048$ we vary DBR with the variation of $q^2$ as well as the new weak phase $\varphi_{sb}$. A noticeable deviation of this parameter from the SM value gives a clue for NP. We find that DBR enhances significantly almost within the whole range of $q^2$ i.e. 14-19 GeV$^2$ and 120-300 degree of $\varphi_{sb}$ for the decay mode $B_s \to \varphi \tau^+ \tau^-$. If we consider the muonic channel represented by Fig. 3(b) then we can see that the enhancement of DBR from SM value through NP are significantly large in low $q^2$ region but in high $q^2$ region this enhancement is quite small. The deviation of $B_s \to \varphi \tau^+ \tau^-$ decay is significantly large compared to $B_s \to \varphi \mu^+ \mu^-$ decay. This provides the lepton flavour non-universality.

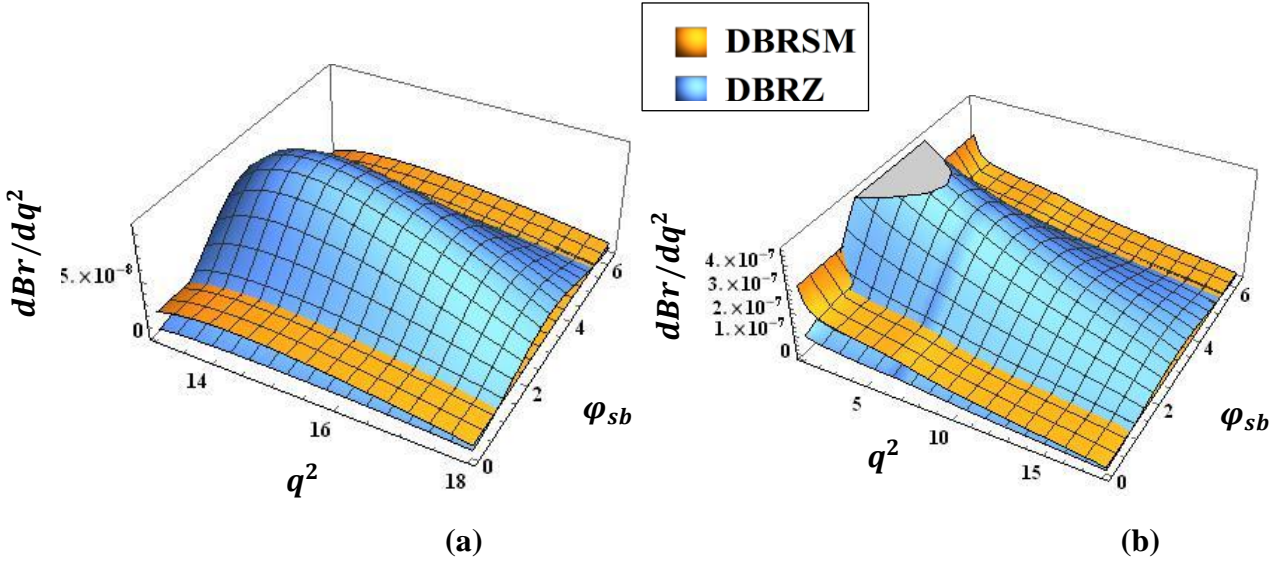

**Fig. 3.** Variation of differential branching ratio $\frac{dBr}{dq^2}$ (DBR) with weak phase $\varphi_{sb}$ (degree) and $q^2$ (GeV)$^2$ for (a) $B_s \to \varphi \tau^+ \tau^-$ and (b) $B_s \to \varphi \mu^+ \mu^-$ decays.

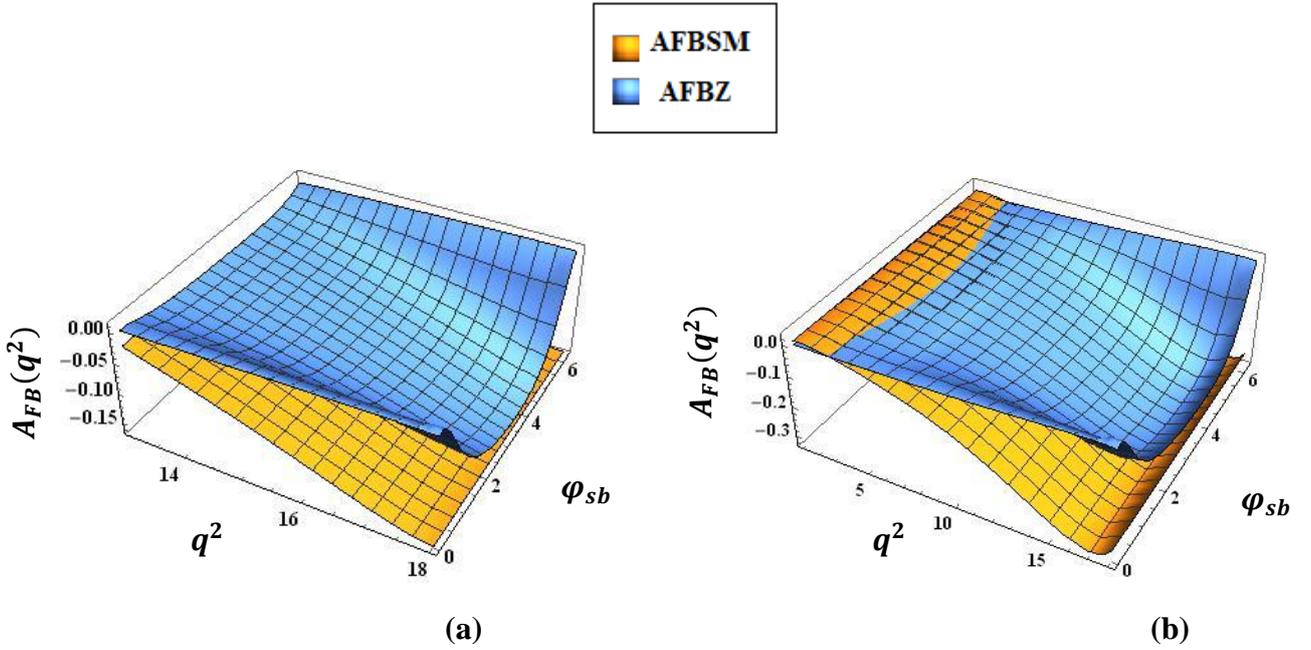

**Fig. 4.** Variation of forward backward asymmetry (FB) $A_{FB}(q^2)$ with weak phase $\varphi_{sb}$ (degree) and $q^2$ (GeV)$^2$ for (a) $B_s \to \varphi \tau^+ \tau^-$ and (b) $B_s \to \varphi \mu^+ \mu^-$ decays.

From Fig. 4(a, b) it is found that the deviation of $A_{FB}$ increases from SM value at very high $q^2$ region. But for the decay channel $B_s \to \varphi \mu^+ \mu^-$ the enhancement of $A_{FB}$ is quite different i.e. $A_{FB}(q^2)$ crosses the SM value at low $q^2$ region and increases significantly from SM value. This indicates lepton flavour non-universality due to different enhancement of $A_{FB}(q^2)$ for $B_s \to \varphi \tau^+ \tau^-$ and $B_s \to \varphi \mu^+ \mu^-$ decay modes.

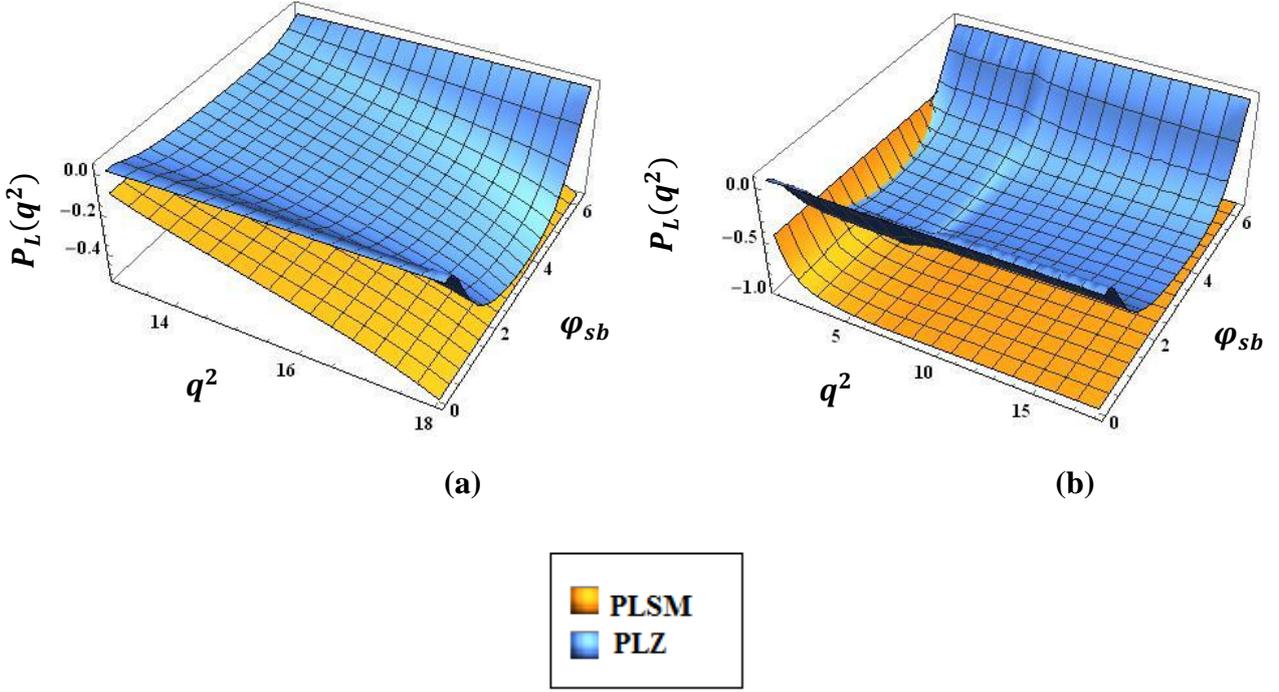

**Fig. 5.** Variation of longitudinal polarization asymmetry $P_L(q^2)$ with weak phase $\varphi_{sb}$ (degree) and $q^2$ (GeV)$^2$ for (a) $B_s \to \varphi \tau^+ \tau^-$ and (b) $B_s \to \varphi \mu^+ \mu^-$ decays.

In Fig. 5, the maximum enhancement of $P_L(q^2)$ is found except the range 120-240 degree of the weak phase $\varphi_{sb}$. These deviations are quite different for decay channels $B_s \to \varphi \tau^+ \tau^-$ and $B_s \to \varphi \mu^+ \mu^-$. In Fig. 5(a), we can also see that this deviation of $P_L(q^2)$ for the decay $B_s \to \varphi \tau^+ \tau^-$ is significantly larger than that of the decays $B_s \to \varphi \mu^+ \mu^-$ shown in Fig. 5(b). This may give the signal of lepton flavour non-universality. Basically this maximum deviation of $P_L(q^2)$ will come due to the significant modifications of $C_9$ and $C_{10}$ in this NP model compared to SM. The most interesting fact is that the value of longitudinal polarization asymmetry is negative in almost the whole region of $q^2$ in SM whereas in NP model it increases quite for $B_s \to \varphi \mu^+ \mu^-$ decay modes. Hence, it gives a signal of NP beyond the SM.

Now, we represent the variations of different kinamatic observables for the decay mode $B^+ \to K^+ l^+ l^-$. In Fig. 6 (a), the maximum enhancement of DBR is found within the range of new weak phase as $120 < \varphi_{sb} < 360$ degree for the tau channel. From Fig. 6 (b), we can see that DBR touches the SM value and crosses it with the increase of weak phase and after a large enhancement it will decrease with the high contribution of $\varphi_{sb}$. We also see that the deviation of DBR from SM value through NP are significantly large in low $q^2$ region but while we consider tau as final state leptons then the noticeable enhancement is observed only within the small portion of total kinematic region i.e. 14-19 GeV$^2$.

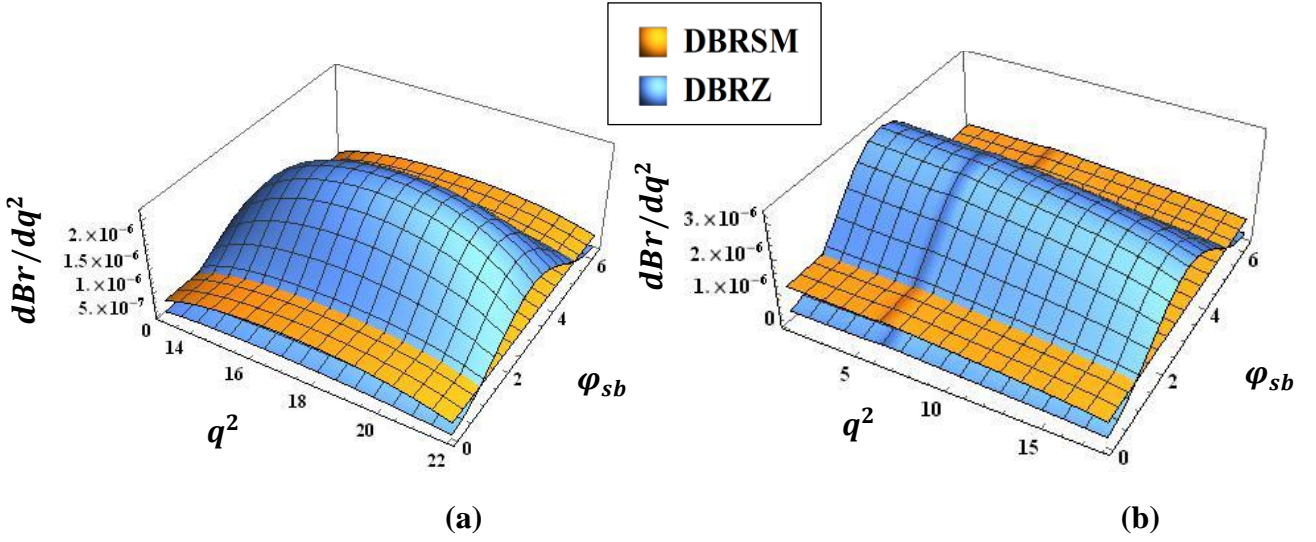

**Fig. 6.** Variation of differential branching ratio $\frac{dBr}{dq^2}$ (DBR) with weak phase $\varphi_{sb}$ (degree) and $q^2$ (GeV)$^2$ for (a) $B^+ \to K^+\tau^+\tau^-$ and (b) $B^+ \to K^+\mu^+\mu^-$ decays.

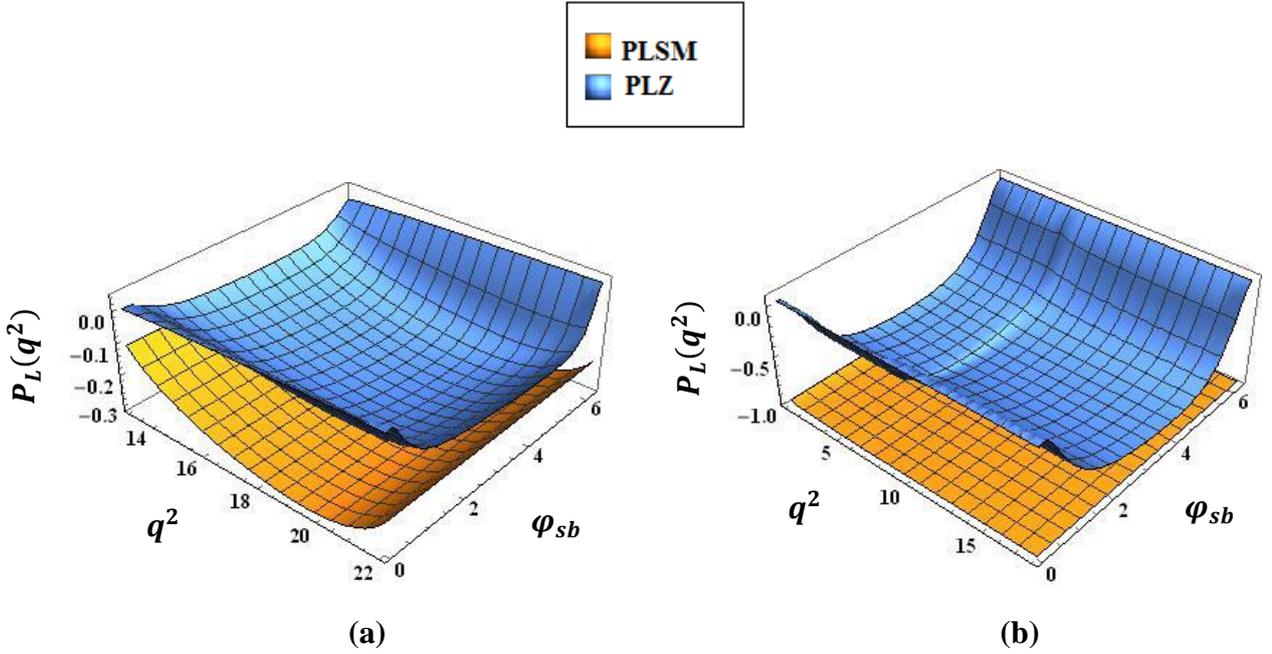

**Fig. 7.** Variation of longitudinal polarization asymmetry $P_L(q^2)$ with weak phase $\varphi_{sb}$ (degree) and $q^2$ (GeV)$^2$ for (a) $B^+ \to K^+\tau^+\tau^-$ and (b) $B^+ \to K^+\mu^+\mu^-$ decays.

For tau channel shown in Fig. 7(a) $P_L(q^2)$ increases sharply from SM value with the increase of $q^2$.

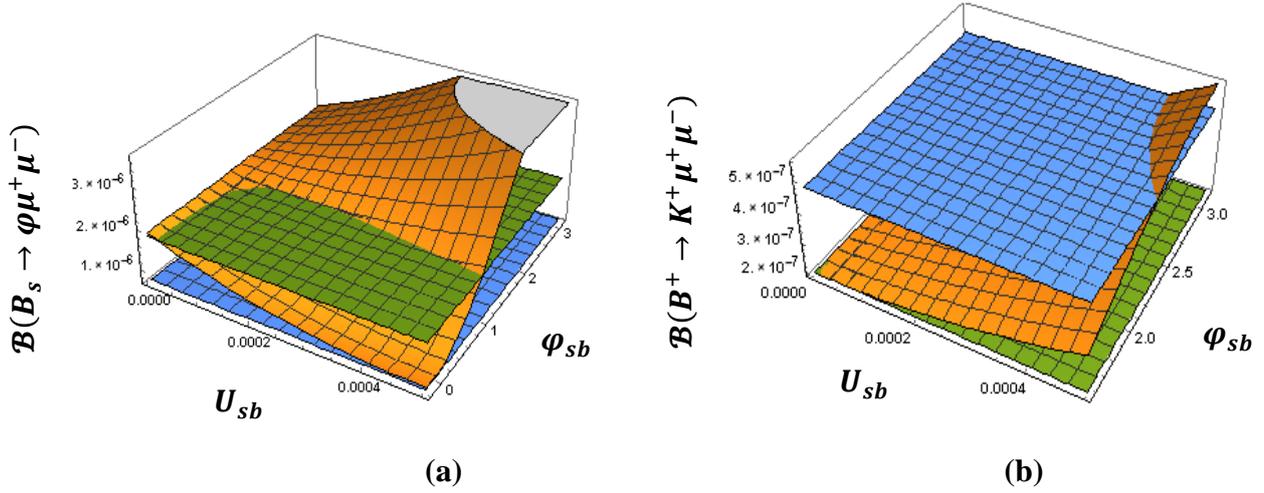

**Fig. 8.** The variation of branching fraction with coupling parameter as $|U_{sb}|$ and weak phase $\varphi_{sb}$ (degree) for (a) $B_s \to \varphi \mu^+ \mu^-$ and (b) $B^+ \to K^+ \mu^+ \mu^-$ decays. Orange plate represents the Z model value, green and blue plates are for SM and experimental values respectively.

**Table 1.** Predicted values of branching fraction of $\mu$ and $\tau$ channel of $B_s \to \varphi l^+ l^-$ and $B^+ \to K^+ l^+ l^-$ decays in SM and NP model. Central values of input parameters as well as form factors are considered for calculation.

| Decay modes | Branching fraction values in the SM | Branching fraction values in Z model | |
|---|---|---|---|
| $\mathcal{B}(B_s \to \varphi \mu^+ \mu^-)$ | $1.80 \times 10^{-6}$ | $7.23 \times 10^{-7}$ | $\varphi_{sb} = 0°$ |
| | | $5.21 \times 10^{-6}$ | $\varphi_{sb} = 180°$ |
| $\mathcal{B}(B_s \to \varphi \tau^+ \tau^-)$ | $1.64 \times 10^{-7}$ | $4.44 \times 10^{-8}$ | $\varphi_{sb} = 0°$ |
| | | $3.69 \times 10^{-7}$ | $\varphi_{sb} = 180°$ |
| $\mathcal{B}(B^+ \to K^+ \mu^+ \mu^-)$ | $1.92 \times 10^{-7}$ | $7.23 \times 10^{-8}$ | $\varphi_{sb} = 0°$ |
| | | $5.34 \times 10^{-7}$ | $\varphi_{sb} = 180°$ |
| $\mathcal{B}(B^+ \to K^+ \tau^+ \tau^-)$ | $5.25 \times 10^{-8}$ | $1.40 \times 10^{-8}$ | $\varphi_{sb} = 0°$ |
| | | $1.66 \times 10^{-7}$ | $\varphi_{sb} = 180°$ |

## 8. Summary and Conclusions

In the recent years, as the experimental evidences of NP are increasing so the semileptonic rare B meson decay channels mediated by $b \to s l^+ l^-$ quark level transition [1-11, 23-32, 93-99] become the fertile ground for theoretical and experimental studies. In this paper, several physical observables for $b \to s l^+ l^-$ mediated decays $B_s \to \varphi l^+ l^-$ and $B^+ \to K^+ l^+ l^-$ in the SM and non-universal Z model are discussed. We have shown the variations of different observables of these decay modes by varying $q^2$ and weak phase $\varphi_{sb}$. Recently, the LHCb Collaboration [100] has found the branching fraction of the decay mode $B_s \to \varphi \mu^+ \mu^-$ as $\mathcal{B}(B_s \to \varphi \mu^+ \mu^-) = (7.07^{+0.64}_{-0.59} \pm 0.17 \pm 0.71) \times 10^{-7}$ over full $q^2$ range. A complete data sets are also found from LHCb experiment [101] to measure the differential branching ratio of $B_s \to \varphi \mu^+ \mu^-$. Another noticeable point is that LHCb collaboration [102] has fitted the differential branching fraction of $B^+ \to K^+ \mu^+ \mu^-$ decay channel by setting the $J/\psi$ and $\psi(2S)$ resonance amplitudes to zero and have obtained $\mathcal{B}(B^+ \to K^+ \mu^+ \mu^-) = (4.37 \pm 0.15 \pm 0.23) \times 10^{-7}$. Whereas recently from Belle experiments [11], it is found that the branching ratio within the whole $q^2$ region becomes $\mathcal{B}(B^+ \to K^+ \mu^+ \mu^-) = (6.24^{+0.65}_{-0.61} \pm 0.31) \times 10^{-7}$. Our predicted values of the branching fractions for $B_s \to \varphi \mu^+ \mu^-$ and $B^+ \to K^+ \mu^+ \mu^-$ decays in the SM and non-universal Z model are shown in Table 1. From this table, it is observed that the SM predicted value for $B(B_s \to \varphi \mu^+ \mu^-)$ is higher than the experimental value whereas for the decay channel $B^+ \to K^+ \mu^+ \mu^-$ the branching ratio value in the SM is lower than the experimental value. In Table 1 and Fig. 8, we observe that our estimated branching ratio value in Z model for $B_s \to \varphi \mu^+ \mu^-$ decay is very close to the experimental value for $\varphi_{sb} = 0°$. Whereas for $B^+ \to K^+ \mu^+ \mu^-$ decay channel our estimated branching ratio value in Z model is close to the experimental value for $\varphi_{sb} = 180°$. Again, the zero position of forward backward asymmetry in the SM plays an important role for searching NP. For the decays channel $B_s \to \varphi \mu^+ \mu^-$ the value of zero crossing for FB asymmetry is approximately $q^2 \approx 2 \text{ GeV}^2$ in Z model. The large deviation of FB asymmetry for both $\tau$ and $\mu$ channel from the value of the SM gives a clear clue for NP. We hope that future data of FB asymmetry of $B_s \to \varphi \mu^+ \mu^-$ decay channel coming from the LHCb and/ or the Belle II detector will help us in finding NP beyond SM. The other optimized parameter is the longitudinal polarization asymmetry which is negative through all over the region of $q^2$ in SM for both $B_s \to \varphi l^+ l^-$ and $B^+ \to K^+ l^+ l^-$ decay modes. From the significant enhancements of the physical observables in non-universal Z model our conclusion is that FCNC mediated Z boson modifies the SM picture and can give the clue for NP beyond the SM. Furthermore, from the different slope of planes we can also find the path of lepton flavour non-universality. We expect that the further investigation of these decay channels at the LHCb and/or at the Belle II in future would be an useful tool to search NP beyond the SM.

**Acknowledgments**

We thank the reviewers for useful comments for the improvement of the manuscript. Nayek and Sahoo are grateful to SERB, DST, Govt. of India for financial support (EMR/2015/000817). Biswas thanks NIT Durgapur for providing fellowship. Maji is thankful to DST, Govt. of India for providing INSPIRE Fellowship (IF160115).

# Appendix A
**Table 2.** Central values of input parameters [88]

| Parameters | Value |
|---|---|
| $m_b$ | 4.18 GeV |
| $m_c$ | 1.275 GeV |
| $M_B$ | 5.366 GeV |
| $M_\varphi$ | 1.019 GeV |
| $|V_{ts}|$ | $39.4 \times 10^{-3}$ |
| $|V_{tb}|$ | 1.019 |
| $\alpha$ | 1/137 |
| $G_F$ | $1.17 \times 10^{-5} \text{GeV}^{-2}$ |
| $m_\tau$ | 1.776 GeV |
| $M_{B^+}$ | 5.279 GeV |
| $M_{K^+}$ | 0.493 GeV |
| $m_\mu$ | $105.65 \times 10^{-3}$ GeV |
| $m_e$ | $548 \times 10^{-6}$ GeV |

# Appendix B: Form factors for $B_s \to \varphi$ transition

The form factors of $B_s \to \varphi$ transition are given as [103, 104].

$$g(q^2) = \frac{V(q^2)}{M_B + M_\varphi} \quad (B1)$$

$$f(q^2) = (M_B + M_\varphi) A_1(q^2) \quad (B2)$$

$$a_+(q^2) = \frac{A_2(q^2)}{M_B + M_\varphi} \quad (B3)$$

$$a_-(q^2) = -\frac{2M_\varphi}{q^2}\left(A_3(q^2) - A_0(q^2)\right) \quad (B4)$$

$$g_+(q^2) = -T_1(q^2) \quad (B5)$$

$$g_-(q^2) = -\frac{M_B^2 - M_\varphi^2}{q^2}\left(T_1(q^2) - T_2(q^2)\right) \quad (B6)$$

$$g_0(q^2) = -\frac{2}{q^2}\left(T_2(q^2) - T_1(q^2) + \frac{q^2}{M_B^2 - M_\varphi^2} T_3(q^2)\right). \quad (B7)$$

The definition of $V(q^2)$, $A_0(q^2)$ and $T_1(q^2)$ can be parameterized as

$$F(q^2) = \frac{r_1}{1 - q^2/m_R^2} + \frac{r_2}{1 - q^2/m_{fit}^2} \quad (B8)$$

Similarly $A_2(q^2)$, $\widetilde{T}_3$ and $A_1(q^2)$, $T_2(q^2)$ are defined by (B9) and (B10) respectively

$$F(q^2) = \frac{r_1}{1 - q^2/m^2} + \frac{r_2}{(1 - q^2/m^2)^2} \tag{B9}$$

$$F(q^2) = \frac{r_2}{1 - q^2/m_{fit}^2} \tag{B10}$$

$T_3(q^2)$ and $A_3(q^2)$ can be parameterized as

$$T_3(q^2) = \frac{M_B^2 - M_\varphi^2}{q^2} \left( \widetilde{T}_3(q^2) - T_2(q^2) \right) \tag{B11}$$

$$A_3(q^2) = \frac{M_B + M_\varphi}{2M_\varphi} A_1(q^2) - \frac{M_B - M_\varphi}{2M_\varphi} A_2(q^2) \tag{B12}$$

### Appendix C: Form factors for $B^+ \to K^+$ transition

The form factors defined for $B^+ \to K^+$ transition are represented as follows [105]

$$f_-(q^2) = \frac{M_B^2 - M_{K^+}^2}{q^2} \left( f_0(q^2) - f_+(q^2) \right) \tag{C1}$$

$$t(q^2) = \frac{f_T(q^2)}{M_B + M_{K^+}}. \tag{C2}$$

Here $f_+(q^2)$, $f_T(q^2)$ and $f_0(q^2)$ and can be expressed by following functions which are represented by equations (C3) and (C4)

$$f_{+(T)}(q^2) = \frac{r_1}{1 - q^2/m_1^2} + \frac{r_2}{(1 - q^2/m_1^2)^2} \tag{C3}$$

$$f_0(q^2) = \frac{r_2}{1 - q^2/m_{fit}^2}. \tag{C4}$$

In the above equations, the parameters used in the fit functions of the form factors are taken from [102, 103] and encapsulated in the following table.

**Table 3.** Input parameters for form factor calculations

| $F(q^2)$ | $F(0)$ | $r_1$ | $m_R^2$ | $r_2$ | $m_{fit}^2(m_1^2)$ |
|---|---|---|---|---|---|
| $f_+(q^2)$ | – | 0.1903 | – | 0.1478 | 29.3 |
| $f_T(q^2)$ | – | 0.1851 | – | 0.1905 | 29.3 |
| $f_0(q^2)$ | – | – | – | 0.3338 | 38.98 |
| $A_1(q^2)$ | 0.311 | – | – | 0.308 | 36.54 |
| $A_2(q^2)$ | 0.234 | −0.054 | – | 0.288 | 48.94 |
| $A_0(q^2)$ | 0.474 | 3.310 | $5.37^2$ | −2.835 | 31.57 |
| $V(q^2)$ | 0.434 | 1.484 | $5.42^2$ | −1.049 | 39.52 |
| $T_1(q^2)$ | 0.349 | 1.303 | $5.42^2$ | −0.954 | 38.28 |
| $T_2(q^2)$ | 0.349 | – | – | 0.349 | 37.21 |
| $\widetilde{T}_3(q^2)$ | 0.349 | 0.027 | – | 0.321 | 45.56 |